\documentclass[letterpaper,twocolumn,pre,aps,showpacs,
floatfix]{revtex4-1}
\usepackage[latin1]{inputenc}
\usepackage{bm}
\usepackage[usenames]{color}
\usepackage{multirow}
\usepackage{amssymb}
\usepackage{amsbsy}
\usepackage{amsmath}
\usepackage{stmaryrd}
\usepackage{graphicx}
\usepackage{epsfig}
\usepackage{placeins}
\usepackage{ulem}
\usepackage{bbm}
\usepackage{braket}
\usepackage[colorlinks,linkcolor=blue,citecolor=blue,urlcolor=blue]{hyperref}
\usepackage{blindtext}

\usepackage{array}

\makeatletter

\begin{document}

\title{Decay of spin-spin correlations in disordered quantum and classical 
spin chains}

\author{Jonas Richter}
\email{jonasrichter@uos.de}
\affiliation{Department of Physics, University of Osnabr\"uck, D-49069 
Osnabr\"uck, Germany}

\author{Dennis Schubert}
\email{dennis.schubert@uos.de}
\affiliation{Department of Physics, University of Osnabr\"uck, D-49069 
Osnabr\"uck, Germany}

\author{Robin Steinigeweg}
\email{rsteinig@uos.de}
\affiliation{Department of Physics, University of Osnabr\"uck, D-49069 
Osnabr\"uck, Germany}

\date{\today}


\begin{abstract}
The real-time dynamics of equal-site correlation functions is studied 
for one-dimensional spin models with quenched disorder. Focusing on infinite 
temperature, we present a comparison between the dynamics of models with 
different quantum numbers $S = 1/2, 1, 3/2$, as well as of chains consisting of 
classical spins.
Based on this comparison as well as by analyzing the statistics of 
energy-level spacings, we show that the putative many-body 
localization transition is shifted to 
considerably stronger values of disorder for increasing $S$. In this context, 
we introduce an effective disorder strength $W_\text{eff}$, which 
provides a mapping between the dynamics for different spin 
quantum 
numbers. For small $W_\text{eff}$, we show that the real-time 
correlations become essentially independent of $S$, and are moreover very well 
captured by the dynamics of classical spins. Especially for $S = 3/2$, the 
agreement between quantum and classical dynamics is remarkably observed 
even for very strong values of disorder. This behavior 
also reflects itself in the
corresponding spectral functions, which are obtained via
a Fourier transform from the time to the frequency domain. As an aside, we also 
comment on the self-averaging properties of the correlation function at 
weak and strong disorder. 
Our work sheds light on the correspondence 
between quantum and classical 
dynamics at high temperatures and extends our understanding of the 
dynamics in disordered spin chains beyond the well-studied case of $S=1/2$.  

\end{abstract}

\maketitle


\section{Introduction}

Noninteracting particles in one and 
two spatial dimensions are localized even 
for arbitrarily small values of disorder \cite{Anderson1958, Abrahams1979}. 
Generalizing this 
well-understood concept of Anderson localization to the presence of 
interactions
has been a major objective of modern condensed matter physics. Based on 
pioneering early works \cite{Gorniyi2005, Basko2006}, and a large number of 
subsequent 
(mostly numerical) studies (see e.g.\ \cite{Oganesyan2007, Pal2010, Imbrie2016, 
Berkelbach2010, Luitz2015, Kjall2014, Bera2015}), many-body localization (MBL) 
is 
nowadays believed to be a generic property in one-dimensional 
short-range lattice models with sufficiently strong randomness 
\cite{Nandkishore2015, Abanin2018}. (For another point of view, 
see \cite{Suntajs2019}.)
The phenomenology of the localized phase is best understood in terms of 
an emergent set of local integrals of motion \cite{Serbyn2013, Huse2014}, 
which, e.g., 
explain the 
slow logarithmic growth of entanglement in time \cite{Znidaric2008, 
Bardarson2012}. Moreover, localized systems violate the eigenstate 
thermalization 
hypothesis \cite{Nandkishore2015, Srednicki1994, Dallesio2016}, which reflects 
itself in the area-law entanglement of eigenstates \cite{Bauer2013}, as well 
as the failure of MBL systems to reach thermal equilibrium at long times. 

Progress in the field of many-body localization has also been fostered by the 
advance of novel 
experimental platforms \cite{Schreiber2015, Orell2019}, which can be very well 
isolated from the 
environment. In particular, such experiments also allow to tackle new 
and exciting questions about the existence of MBL, e.g., in dimensions larger 
than one \cite{Choi2016}, in the presence of long-range interactions 
\cite{Smith2016}, or in 
systems which are weakly coupled to a thermal bath \cite{Lueschen2017}. 

More recently, there has also been much effort to put the phenomenon of MBL 
into a broader perspective and understand the dynamics of models, which might 
not be strictly localized but nevertheless exhibit anomalously small relaxation 
rates and very long equilibration times \cite{Gopalakrishnan2019}.    
Examples include, e.g., disordered models with power-law interactions 
\cite{Nandkishore2017}, systems with two particle species and large mass 
ratio \cite{Grover2014, Schiulaz2014, Yao2016, Sirker2019}, as well as Hubbard 
models where the disorder only acts on the charge degrees of freedom but not on 
the spin \cite{Prelovsek2016, Kozarzewski2018}.   

In this context, the present work scrutinizes the dynamics and the fate of 
MBL for a class of 
disordered models which have received less attention so far.
In particular, while a plethora of works has explored many-body 
localization in disordered spin-$1/2$ systems, 
much less is known about the 
effect of disorder for larger spin quantum numbers 
$S>1/2$ \cite{Richter2019}. Since genuine MBL is believed to be a pure
quantum phenomenon, it 
is 
an intriguing and open question to what extent localization can occur
upon 
increasing $S$, where the quantum spins are supposed to 
become more and more akin to their classical counterparts.
In this work, we shed light onto this question by comparing the 
infinite-temperature dynamics 
of equal-site correlation functions for disordered classical and quantum spin 
chains with $S = 1/2,1,3/2$.
Based on this comparison as well as by analyzing the statistics of
energy-level spacings, we show that the putative many-body localization 
transition is shifted to
considerably stronger values of disorder for increasing $S$. 
Furthermore, introducing an effective disorder $W_\text{eff}$, we 
find a mapping between the dynamics
for different quantum numbers $S$. For small $W_\text{eff}$, we 
show that the real-time correlations 
become essentially independent of $S$, and are moreover very well captured by 
the dynamics of classical spins.
Especially for $S = 3/2$, the agreement between quantum and classical 
dynamics 
is remarkably observed
even for very strong values of disorder. 

This paper is structured as follows. In Secs.\ \ref{Sec::Model}, 
\ref{Sec::Observ}, and \ref{Sec::Num}, we introduce the models and observables 
which are studied in this work, and explain our  
numerical approach. In Sec.\ \ref{Sec::Results}, we then present our numerical 
results for the dynamics in clean and disordered quantum and classical spin 
chains. 
Finally, we summarize and discuss our findings in 
Sec.\ \ref{Sec::Discussion}.  


\section{Setup}\label{Sec::Setup}

\subsection{Model}\label{Sec::Model}

We consider the one-dimensional Heisenberg model with quenched disorder and 
periodic boundary conditions, described by the Hamiltonian 
\begin{equation}\label{EqHamil}
  {\cal H} = J\sum_{l=1}^L \left( {\bf S}_l {\bf S}_{l+1} + h_l S_l^z \right)\ 
, 
\end{equation}
where $L$ denotes the number 
of lattice sites, $J = 1$ is the antiferromagnetic 
exchange constant, and the on-site magnetic fields $h_l$ are randomly 
drawn from a uniform distribution $h_l \in [-W,W]$, with $W$ setting 
the magnitude of disorder.
Moreover, the ${\bf S}_l = (S_l^x,S_l^y,S_l^z)$ are either quantum spin-$S$ 
operators with $S = 1/2,1,3/2$ or, in the classical case, real three-component 
vectors of unit length. Note that ${\cal H}$ conserves the 
total magnetization, i.e., $[\sum_l S_l^z, {\cal H}] = 0$, for all values of 
$W$ and $S$. Note further that for the 
particular case of $S = 1/2$, 
this model is equivalent via a Jordan-Wigner transformation to interacting 
spinless fermions hopping in a random 
potential. Moreover, for $S = 1/2$ and $W 
= 0$, ${\cal H}$ is integrable in terms of the Bethe ansatz. 

It is instructive to briefly consider two limiting cases of the model 
\eqref{EqHamil}, namely (i) the most quantum case $S = 1/2$ and (ii) 
the case of classical spins. 
On the one hand, for $S = 1/2$, the Hamiltonian \eqref{EqHamil} has 
become a canonical model to study the phenomenon of many-body localization. 
Specifically, the spin-$1/2$ model is believed to undergo a transition from an 
ergodic phase into a many-body localized regime above a 
critical disorder strength $W_c \approx 3.5$ (see e.g.\ \cite{Pal2010, 
Luitz2015}), although also larger 
values have been suggested \cite{Devakul2015, Doggen2018}. 
On the other hand, also in the case of 
classical spins, strong disorder has been shown to cause a drastic reduction 
of transport coefficients and anomalously slow relaxation 
\cite{Oganesyan2009, Jencic2015}.  
The occurrence of genuine MBL, however, is not expected for classical spin 
models. This can be understood for instance from a rare-region argument, 
where small parts of the chain are only weakly disordered and exhibit chaotic 
dynamics, 
which eventually causes thermalization of the full system at sufficiently long 
time scales 
\cite{Gopalakrishnan2019}.  

Not least due to the higher computational requirements, much less is known 
about the dynamics and the putative MBL transition for the Hamiltonian 
\eqref{EqHamil} between these two limiting cases, i.e., for larger 
spin quantum numbers $S > 1/2$. The present work  
attempts to bridge this gap and sheds light on the dynamics in disordered 
spin chains with $S = 1, 3/2$.   

\subsection{Equal-site correlation function}\label{Sec::Observ}

As a central quantity of interest in this paper, we study the dynamics 
of the disorder-averaged (denoted by the overbar) equal-site correlation 
function 
\begin{equation}\label{Eq::Correl}
 \overline{C(t)} = \overline{\langle S_l^z(t) S_l^z \rangle}\ , 
\end{equation}
for the disordered spin model \eqref{EqHamil} at formally infinite temperature. 
Note that due to periodic boundary 
conditions, the specific site $l$ to calculate $\overline{C(t)}$ is arbitrary 
(upon 
averaging the bare $C(t)$ over sufficiently many disorder realizations, cf.\ 
Sec.\ 
\ref{Sec::Av} below). 

In the quantum case, $C(t)$ is given by 
\begin{equation}\label{Eq::Trace}
 C(t) = \frac{\text{Tr}[S_l^z(t) S_l^z]}{d}\ , 
\end{equation}
where $d = (2S+1)^L$ is the dimension of the Hilbert space, and $S_l^z(t) 
= e^{i{\cal H}t} S_l^z e^{-i{\cal H}t}$ is the time-evolved operator in 
the Heisenberg picture.
Moreover, in the case of classical mechanics, $\langle \cdot \rangle$ 
denotes the ensemble average over classical trajectories,
where for every trajectory all the 
${\bf S}_l$ are randomly initialized at $t = 0$ (within the constraint of 
$|{\bf S}_l| = 1$), and evolve in time according to the Hamiltonian equations 
of motion
\begin{equation}\label{Eq::HEM}
\dot{{\bf S}}_l = \frac{\partial {\cal 
H}}{\partial {\bf S}_l} \times {\bf S}_l\ .
\end{equation} 

In addition to $\overline{C(t)}$, we also study the corresponding spectral 
function 
$\overline{C(\omega)}$, which follows from a Fourier transform of the real-time 
data 
according to 
\begin{equation}\label{Eq::CW}
 \overline{C(\omega)} = \int_{-\infty}^\infty \overline{C(t)}e^{i\omega t} 
\text{d}t\ , 
\end{equation}
and which is relevant to, e.g., the relaxation rate probed in nuclear magnetic 
resonance experiments \cite{Herbrych2013}. Note that, since the real-time 
dynamics of $\overline{C(t)}$ can in practice only be evaluated up to a finite 
cutoff time 
$t_\text{max} < \infty$, the Fourier transform \eqref{Eq::CW} yields the 
spectral function $\overline{C(\omega)}$ with a finite frequency resolution 
$\delta \omega 
= \pi/t_\text{max}$.

\subsubsection{Decay of the correlation function}

The autocorrelation function $\overline{C(t)}$ in Eq.\ \eqref{Eq::Correl} can 
be 
interpreted as the ``return probability'' of a single spin excitation which 
is created at some lattice site $l$, and spreads within the system with 
respect to an infinite-temperature many-body background.

On the one hand, in the ergodic phase, the dynamics of $\overline{C(t)}$ can 
be understood as follows. After a quick initial decay starting from 
\begin{equation}\label{Eq::Initial}
 \overline{C(0)} = 
\frac{S(S+1)}{3}\ , 
\end{equation}
$\overline{C(t)}$ is eventually expected to display some 
slower hydrodynamic behavior (since the total $S^z$ is conserved), 
which reflects itself in a power-law decay
\begin{equation}
\overline{C(t)} \propto t^{-\alpha}\ . 
\end{equation}
For one-dimensional systems, $\alpha = 1/2$ would here correspond to 
conventional diffusive transport. In this context, let us note that for the 
spin-$1/2$ version of Eq.\ \eqref{EqHamil}, there is numerical evidence 
for the existence of a subdiffusive phase in the regime of low to intermediate 
disorder below the localization transition 
\cite{Agarwal2015, Gopalakrishnan2015, 
Znidaric2016, Luitz2017}. This subdiffusive 
regime manifests itself, e.g., in an anomalously slow relaxation of the 
autocorrelation function $\overline{C(t)}$ with exponent $\alpha < 0.5$, as 
well 
as sublinear buildup of entanglement \cite{BarLev2015, Luitz2016, 
Luitz2016_2, Khait2016, Bera2017}. Eventually, for even longer 
times, the 
spin excitation has explored the entire system (assuming a finite chain length 
$L$), and $\overline{C(t)}$ will saturate to a constant long-time value which 
scales as 
$\propto 1/L$ \cite{Richter2018}.

On the other hand, in the localized regime, the decay of 
the autocorrelation function is significantly slower. In particular, 
$\overline{C(t)}$ saturates to a nonzero value at long times (or to a value 
$>1/L$ in a 
finite system). In this sense, the long-time value of $\overline{C(t)}$ can be 
used as an 
order parameter for the MBL transition \cite{Luitz2017}.  


\subsection{Numerical approach}\label{Sec::Num}

\subsubsection{Quantum dynamics}\label{Sec::NumQD}

The numerical evaluation of the quantum expectation value 
\eqref{Eq::Trace} is 
a challenging 
task, not 
least due to the exponential growth of the Hilbert space and the 
necessity to study long time scales.
We here tackle this numerical challenge by using an efficient 
pure-state approach based on the concept of dynamical quantum typicality (DQT) 
\cite{Hams2000, Iitaka2003, Popescu2006, Goldstein2006, 
Reimann2007, Bartsch2009, Sugiura2013, Elsayed2013, Monnai2014, 
Steinigeweg2014, Reimann2018}. 
Within this concept, the equal-site correlation function $C(t)$ in 
Eq.\ \eqref{Eq::Trace} can 
be 
written as the expectation value within a 
single pure state \cite{Richter2019, Richter2018_2, Richter2019_3}, 
\begin{equation}\label{Eq::Typ}
 C(t) = \bra{\psi(t)}S_l^{z}\ket{\psi(t)} + 
\epsilon\ , 
\end{equation}
where $\ket{\psi(0)}$ is prepared according to 
\begin{equation}\label{Eq::Sqroot}
 \ket{\psi(0)} = \frac{\sqrt{S_l^z + 
S}\ket{\varphi}}{\sqrt{\braket{\varphi|\varphi}}}\ ,\ \ (S= \tfrac{1}{2}, 1, 
\tfrac{3}{2})\ , 
\end{equation}
and the reference pure state $\ket{\varphi} = \sum_{k=1}^{d} c_k \ket{k}$ is 
drawn at random from the full Hilbert space according to the unitary invariant 
Haar measure \cite{Bartsch2009}. In practice, the states $\ket{k}$ here denote 
the product 
basis of the local $S^z$ projections (Ising basis), and the complex 
coefficients $c_k$ are randomly drawn from a Gaussian distribution with zero 
mean and unit variance. As an aside, let us note that the application 
of the square root in Eq.\ \eqref{Eq::Sqroot} becomes straightforward since 
$S_l^z + S$ is nonnegative and diagonal in our 
working basis. Importantly, the statistical error $\epsilon = 
\epsilon(\ket{\varphi})$ in Eq.\ 
\eqref{Eq::Typ} scales as 
$\epsilon \propto 1/\sqrt{d}$ \cite{Hams2000, Sugiura2013}, and becomes very 
small even for moderate 
system sizes $L$ (especially for $S > 1/2$). (We demonstrate the smallness of 
statistical errors further below in Sec.\ \ref{Sec::SA}.)
Moreover, thanks to the 
sparse matrix structure of generic few-body 
operators, the time evolution of the pure state $\ket{\psi(t)}$ can be 
efficiently generated 
by iteratively solving the real-time Schr\"odinger equation,
\begin{equation}
\ket{\psi(t+\delta 
t)} = e        
^{-i{\cal H}\delta t} \ket{\psi(t)}\ , 
\end{equation}
with a small discrete time step $\delta t$ by means of, e.g., Runge-Kutta 
(RK) schemes \cite{Elsayed2013, Steinigeweg2014}, Krylov subspace techniques 
\cite{Varma2017}, 
Trotter product formulas \cite{DeRaedt2006}, or Chebyshev expansions 
\cite{Dobrovitski2003, Weisse2006}. Due to this fact, we can treat system sizes 
beyond the range of standard exact 
diagonalization (ED) (here up to $L = 14$ for $S = 3/2$, which is equivalent to 
$L = 28$ for $S = 1/2$), and study comparatively long time scales.  

Eventually, let us emphasize that DQT is independent of the validity of the 
ETH and solely relies on the largeness of the Hilbert space. Thus, DQT also 
allows for accurate calculations in strongly disordered models which undergo a 
MBL transition \cite{Steinigeweg2016}.

Independent of the typicality relation in Eq.\ \eqref{Eq::Typ},
each side of this relation can be interpreted as a certain type of imperfect
``echo protocol'' (see also \cite{Schmitt2018}). To make this a bit more transparent, the
r.h.s.\ of Eq.\
\eqref{Eq::Typ} can be slightly rewritten as
\begin{equation}
\langle \psi(0) | e^{i {\cal H} t} \, S_l^z \, e^{-i {\cal H} t} \, \psi(0) \rangle
\, ,
\end{equation}
i.e., after evolving the out-of-equilibrium initial state $\ket{\psi(0)}$
for some time $t$, a ``perturbation'' is applied in form of the operator
$S_l^z$. Subsequently, the perturbed state is evolved backwards in time and
the overlap between the resulting state $e^{i{\cal H}t} \, S_l^z \, 
e^{-i{\cal H}t} \ket{\psi(0)}$ and the initial state $\ket{\psi(0)}$ is
measured.

\subsubsection{Classical dynamics}

In contrast to the quantum case, the memory requirements for the 
classical spin chain do not scale exponentially, but only linearly in 
system size. 
Therefore, the Hamiltonian equations of motion \eqref{Eq::HEM} can be solved for 
huge systems 
and long times with significantly less computational 
resources. Specifically, we here employ a 
fourth-order RK scheme, where $\delta t = 0.01$ is chosen  
short enough to conserve the total energy and total magnetization to very 
high accuracy \cite{Steinigeweg2012, Das2018}. For a proper comparison between 
quantum 
and classical dynamics, 
however, we here present data mostly for system sizes $L$, which can also be 
treated quantum mechanically. 

\subsubsection{Averaging}\label{Sec::Av}

As it is standard in context of disordered systems, the bare 
correlations $C(t)$ have to be averaged over $N_h$ independent 
realizations of the random magnetic fields $h_l$ in order to obtain reliable 
results, 
\begin{equation}
 \overline{C(t)} = \frac{1}{N_h}\sum_{h=1}^{N_h} C(t)\ .
\end{equation}
Accordingly, the results for $\overline{C(t)}$ presented in this 
paper 
have to be understood as the average from $N_h = 100 - 1000$ random 
realizations, depending on the system size $L$ and the spin 
quantum number $S$. (Note that in the case of classical spins, the ensemble 
average $C(t)$ for a given disorder realization is obtained from approximately 
$10^3$ independent trajectories.)

Moreover, in Sec.\ \ref{Sec::SA}, we are concerned with the question whether or 
not $C(t)$ is a self-averaging quantity. Here, self-averaging refers to the 
fact that for larger and larger systems, fewer and fewer disorder realizations 
are needed to correctly represent the whole statistical ensemble. To this end, 
we introduce the relative variance $R(t)$ of sample-to-sample 
fluctuations \cite{Schiulaz2019}, 
\begin{equation}\label{Eq::Rt}
 R(t) = \frac{\overline{C(t)^2} - \overline{C(t)}^2}{\overline{C(t)}^2}\ . 
\end{equation}
We call $C(t)$ self-averaging if $R(t)$ decreases upon increasing the system 
size, i.e., $R(t) \propto L^{-\nu}$ with $0 < \nu \leq 1$, where $\nu = 1$ 
would correspond to strong self-averaging.  

As another probe of self-averaging, we also consider the log-averaged 
correlation function $\overline{C_\text{log}(t)}$ \cite{Serbyn2017}, defined as 
\begin{equation}\label{Eq::Ctlog}
  \overline{C_\text{log}(t)} = \exp\left(\frac{1}{N_h}\sum_{h = 1}^{N_h} \ln 
\left[ C(t) \right] \right)\ .
\end{equation}
Naturally, the corresponding log-averaged spectral function 
$\overline{C_\text{log}(\omega)}$ follows from Eq.\ \eqref{Eq::CW} upon 
replacing $\overline{C(t)}$ by $\overline{C_\text{log}(t)}$.
It is easy to see that the log-average $\overline{C_\text{log}(t)}$ is in 
fact equivalent to the geometric mean (whereas  
$\overline{C(t)}$ is the ``standard'' arithmetic mean), which is also the 
reason why $\overline{C_\text{log}(t)}$ is often referred to as {\it 
typical} correlation \cite{Agarwal2015, Serbyn2017}. 


\section{Results}\label{Sec::Results}

Let us now present our numerical results. In Sec.\ \ref{Sec::Results_Clean}, we 
begin by studying the clean Heisenberg chain ($W = 0$) for $S = 1/2,1,3/2$ and
classical spins, which serves as a convenient
starting point for the actual discussion of disordered models in Sec.\ 
\ref{Sec::Results_Disordered}. Note that, in order to simplify 
the notation, we drop the overbar for the averaged quantities in the 
following. Thus, whenever we write $C(t)$ or $C_\text{log}(t)$, we implicitly
refer 
to $\overline{C(t)}$ or $\overline{C_\text{log}(t)}$.

\subsection{Clean model}\label{Sec::Results_Clean}

\begin{figure}[tb]
 \centering
 \includegraphics[width=0.85\columnwidth]{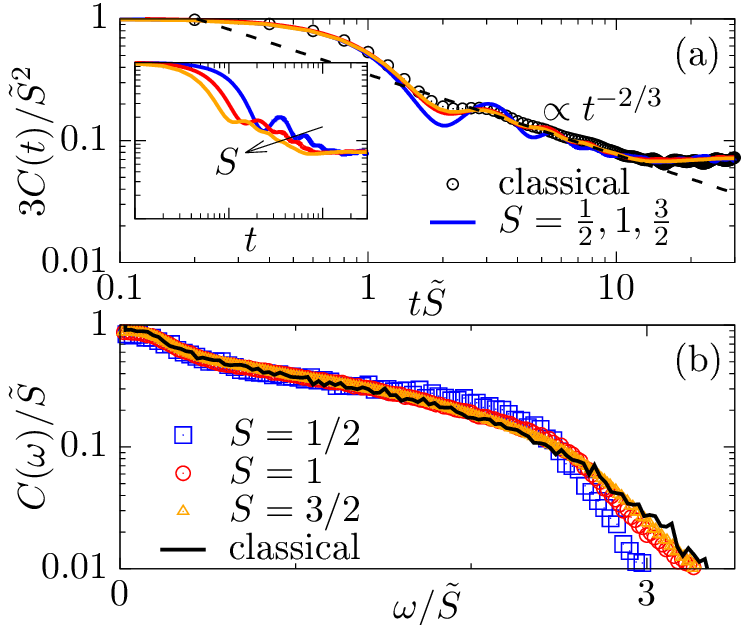}
 \caption{(Color online) (a) 
Equal-site correlation function $3C(t)/\widetilde{S}^2$ for spin quantum 
numbers 
$S = 1/2,1,3/2$ (curves) and classical spins (symbols) versus rescaled time
$t\widetilde{S}$. The dashed line indicates a power-law decay $\propto 
t^{-2/3}$. The inset shows the same $S = 1/2,1,3/2$ data, but now versus 
the bare time 
$t$. (b) Corresponding spectral function $C(\omega)/\widetilde{S}$ versus 
rescaled 
frequency $\omega/\widetilde{S}$. We have $L = 14$ in all cases.}
 \label{Fig1}
\end{figure}

In Fig.\ \ref{Fig1}~(a), we show the equal-site correlation function 
$C(t)$ at a fixed system 
size $L = 14$, both for classical spins as well as quantum spins
with $S=1/2,1,3/2$ at vanishing disorder $W = 0$. In order to account for 
the different quantum 
numbers $S$, the time 
axis is renormalized according to 
\cite{Steinigeweg2010}
\begin{equation}
 t \to t\widetilde{S}\ ,\quad \text{with}\ \widetilde{S} = \sqrt{S(S+1)}\ .
\end{equation}
(Note that $\widetilde{S} = 1$ in the case of classical spins.) Moreover, the 
curves are normalized by their initial value $C(0) = \widetilde{S}^2/3$, cf.\ 
Eq.\ \eqref{Eq::Initial}. 
In particular, we find that this rescaling yields a convincing data 
collapse of $C(t)$, and we can identify an intermediate time window $2 \lesssim 
t\widetilde{S} \lesssim 10$, where $C(t)$ decays as a power law, 
$C(t) \propto t^{-\alpha}$, with exponent $\alpha \approx 2/3$ for all curves 
shown here. On the one hand, for the integrable $S = 1/2$ chain, this is 
consistent with other studies 
\cite{Ljubotina2017, Gopalakrishnan2019_2, Richter2019_2, Capponi2019}, which 
report 
that spin 
transport in the isotropic Heisenberg chain is described by 
the Kardar-Parisi-Zhang universality class \cite{Ljubotina2019, Weiner2019}. On 
the other hand, for the nonintegrable $S = 1,3/2$ models, the situation is less 
settled. While it has been put forward that superdiffusion might persist 
despite the absence of integrability \cite{Richter2019, DeNardis2019}, it has 
also been argued 
that $\alpha$ will eventually approach its diffusive value 
$\alpha = 1/2$ asymptotically for larger system sizes and longer time scales 
\cite{Dupont2019}. 
In this context, let us note that even for the classical spin chain, the  
unambiguous detection of diffusion is known to be a delicate problem 
\cite{Mueller1988, Gerling1989, Gerling1990, Alcantara1992, Boehm1993}.

In Fig.\ \ref{Fig1}~(a), we moreover find that the spin-$1/2$ curve exhibits 
some additional oscillations at 
intermediate times, whereas the curves for larger $S = 1, 3/2$ are almost 
indistinguishable from the dynamics of classical spins. We trace these 
oscillations back to (i) enhanced 
quantum fluctuations at $S = 1/2$, and (ii) specific properties of the 
model under consideration. For additional numerical results of $C(t)$ in other
disorder-free models, we refer to Appendix 
\ref{Sec::App1}. Interestingly, breaking the integrability of the $S = 1/2$ 
model, e.g., by means of a next-nearest 
neighbor interaction, not necessarily leads to a further improvement of the 
data collapse. 

In addition to the real-time data, the corresponding 
spectral functions $C(\omega)$ are shown in Fig.\ \ref{Fig1}~(b). Also in this 
case, we find that a proper rescaling of the 
energy scale, $\omega \to \omega/\widetilde{S}$, 
results in a convincing data 
collapse. In fact, notable deviations can only be 
observed at large frequencies $\omega/\widetilde{S} \gtrsim 2.5$, where  
$C(\omega)$ decays exponentially \cite{Abanin2015}, and this decay turns out to 
be slightly faster for $S = 1/2$.

The data presented in Fig.\ \ref{Fig1} already exemplify one important result 
of the present paper. Namely, there exist models and observables where quantum 
dynamics (i) becomes almost independent of the specific quantum number 
$S$ and (ii) can be well captured by the dynamics of classical spins. 
Admittedly, this fact might not be entirely surprising since we are dealing 
with dynamics at formally infinite 
temperature where quantum effects are less pronounced. Nevertheless, it is 
still a remarkable result that the dynamics of classical spins not only 
qualitatively, but even quantitatively reproduces quantum dynamics for $S = 
1, 3/2$. As shown below, this agreement between quantum and classical 
dynamics becomes even more interesting when studying the dynamics in models 
with finite disorder $W > 0$.

\subsection{Disordered model} \label{Sec::Results_Disordered}

\subsubsection{Level-spacing distribution}

\begin{figure}[tb]
 \centering 
 \includegraphics[width = 0.95\columnwidth]{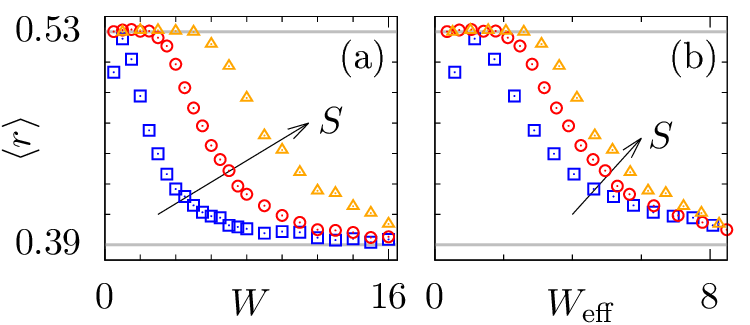}
 \caption{(Color online) Mean ratio $\langle r 
\rangle$ of adjacent level spacings 
for the model 
\eqref{EqHamil} with $S = 1/2,1,3/2$ (arrow) versus 
(a) 
bare disorder strength $W$ and (b) effective disorder $W_\text{eff} = 
W/\widetilde{S}$. The gray lines indicate the value for Wigner-Dyson ($\langle 
r 
\rangle \approx 0.53$) and Poisson distribution ($\langle r 
\rangle \approx 0.39$). We have $L = 8$ in all 
cases.}
 \label{Fig2}
\end{figure}

Before studying dynamics for $W > 0$, a first orientation for the putative MBL 
transition in the model \eqref{EqHamil} with $S \geq 1/2$ 
can be obtained from the mean ratio $\langle r 
\rangle$ of adjacent level spacings,  
\begin{equation}
 \langle r \rangle = \left\langle \frac{\text{min}\lbrace 
 \delta_n,\delta_{n+1}\rbrace}{\text{max}\lbrace \delta_n,\delta_{n+1} 
\rbrace} 
 \right\rangle\ , 
 \end{equation}
 where $\delta_n = |E_{n+1} - E_n|$ with $E_n$ being the energy-ordered 
eigenvalues of ${\cal H}$, and the brackets denote both, averaging 
over approximately $1/3$ of the levels in the center of the 
spectrum, and over sufficiently many independent realizations of the random 
$h_l$. For an ergodic 
system, the 
level statistics is supposed to follow 
a Wigner-Dyson distribution with $\langle r \rangle \approx 0.53$, whereas the 
occurrence of MBL at strong disorder reflects itself by the onset of Poissonian 
level statistics with $\langle r \rangle \approx 0.39$ \cite{Pal2010}. 

Our numerical results for $\langle r \rangle$ are summarized in Fig.\ 
\ref{Fig2}, where we restrict ourselves to a fixed 
system size $L = 8$ and eigenstates in the zero-magnetization subsector. (Note 
that ED becomes very costly for $S = 3/2$ and $L > 8$.)
As can be seen in Fig.\ \ref{Fig2}~(a), the ratio $\langle r \rangle$ 
interpolates between $\langle r \rangle = 0.53$ and $\langle r \rangle = 
0.39$ for all values of $S = 
1/2,1,3/2$ upon increasing the disorder 
strength $W$. (The small dip of 
$\langle r \rangle$ for $S = 1/2$ and $W = 0.5$ can be explained due to 
the vicinity of the integrable point at $W = 0$.) Importantly, however, we 
find that the transition of $\langle r \rangle$ from Wigner-Dyson towards 
Poissonian level statistics occurs at larger and larger values of $W$ if the 
spin quantum number $S$ is increased. This result suggests that 
a transition to a many-body localized regime in spin chains with  
$S > 1/2$, if existent at all, probably requires a stronger and stronger 
critical disorder $W_c$ (compared to $W_c \approx 3.5$ for $S = 
1/2$). This is an important result of the 
present paper. 
\begin{table}[b]
\caption{Disorder values used in the 
simulations to obtain the data shown in Fig.\ \ref{Fig3} 
and \ref{Fig4}. As a point of reference, we set the bare disorder for the 
spin-$1/2$ model to $W = 
0.5,1,2,5$, yielding the corresponding effective disorder values $W_\text{eff} 
= W/\sqrt{3/4}$, cf.\ Eq.\ \eqref{Eq::Wtilde}. The bare disorder values for $S 
= 1,3/2$ consequently follow from $W = 
W_\text{eff}\widetilde{S}$. Note that the values are rounded to two decimal 
places.}
\label{Tab1}
 \begin{tabular}{|p{1.5cm}|p{1.5cm}|p{1.5cm}||c||}
 \hline
 \multicolumn{3}{|c||}{bare disorder $W$} &  \multirow{2}{2cm}{ 
\centering effective disorder $W_\text{eff}$} 
\\ \cline{1-3} \cline{1-3}
  \centering $S = 1/2$ &  \centering $S = 1$ &  \centering $S = 3/2$ &  \\ 
\hline \hline
  
 \centering 0.5 & \centering 0.82 & \centering 1.12 & 0.58 \\ 
\hline 
   \centering 1 &  \centering 1.63 &  \centering 2.24  &   1.15 \\ \hline
   \centering 2 &  \centering 3.27 &  \centering 4.47  &   2.31 \\ \hline
   \centering 5 &  \centering 8.16 &  \centering 11.18 &   5.77 \\ \hline
 \end{tabular}
\end{table}

In view of the successful rescaling found in the 
context of Fig.\ \ref{Fig1}, let us now introduce an effective 
disorder strength $W_\text{eff}$ according to, 
\begin{equation}\label{Eq::Wtilde}
 W_\text{eff} = W/\widetilde{S} = W/\sqrt{S(S+1)}\ . 
\end{equation}
Plotting $\langle r \rangle$ 
versus $W_\text{eff}$ in Fig.\ \ref{Fig2}~(b), we in fact observe a reasonable 
data collapse for small $W_\text{eff} \approx 1$ and large 
$W_\text{eff} \approx 
8$. In the intermediate regime $W_\text{eff} \approx 4$, 
however, this collapse clearly breaks down and $\langle r \rangle$ is still 
generally larger for larger $S$. 
Given the facts that (i) $\langle r \rangle$ naturally depends on the 
system size $L$ \cite{Pal2010} (see also Appendix \ref{Sec::FSLS} for additional 
data), and (ii) the proper analysis of finite-size data around the 
critical disorder is known to be very delicate \cite{Abanin2019}, it is 
difficult to estimate whether or not this behavior persists in the 
thermodynamic limit $L \to \infty$. 
Nonetheless, the effective disorder $W_\text{eff}$ in Eq.\ \eqref{Eq::Wtilde} 
actually turns out to be very useful when comparing $C(t)$ and $C(\omega)$ for 
different quantum numbers $S$ below.

\subsubsection{Dynamics}\label{Sec::DyDisorder}

Let us now proceed to the actual discussion of $C(t)$ and $C(\omega)$ in  
disordered spin 
chains. 
In particular, it is instructive to compare the dynamics for different quantum 
numbers $S$, while using the 
same value of the effective disorder 
$W_\text{eff}$ for each $S$. Since the MBL transition in the $S = 1/2$ chain is 
well-studied, this model serves as a point of reference and we set the 
values of the bare disorder $W$ for the $S = 1/2$ chain to $W = 0.5,1,2$ and 
$W = 5$. In 
view of the putative critical disorder $W_c \approx 3.5$, we thus study the 
dynamics of $C(t)$ and $C(\omega)$ both in the ergodic and in the localized 
regime of the spin-$1/2$ model. Moreover, given these values of $W$ for $S = 
1/2$, we consequently obtain effective disorder values 
$W_\text{eff} = W / \sqrt{3/4} \approx 0.58, 1.15, 2.31, 5.77$. The 
corresponding bare disorder values $W$ for $S = 
1, 3/2$ then directly follow from $W = W_\text{eff} \sqrt{S(S+1)}$ and are 
summarized in Table \ref{Tab1}. 
\begin{figure}[t]
 \centering
 \includegraphics[width=1\columnwidth]{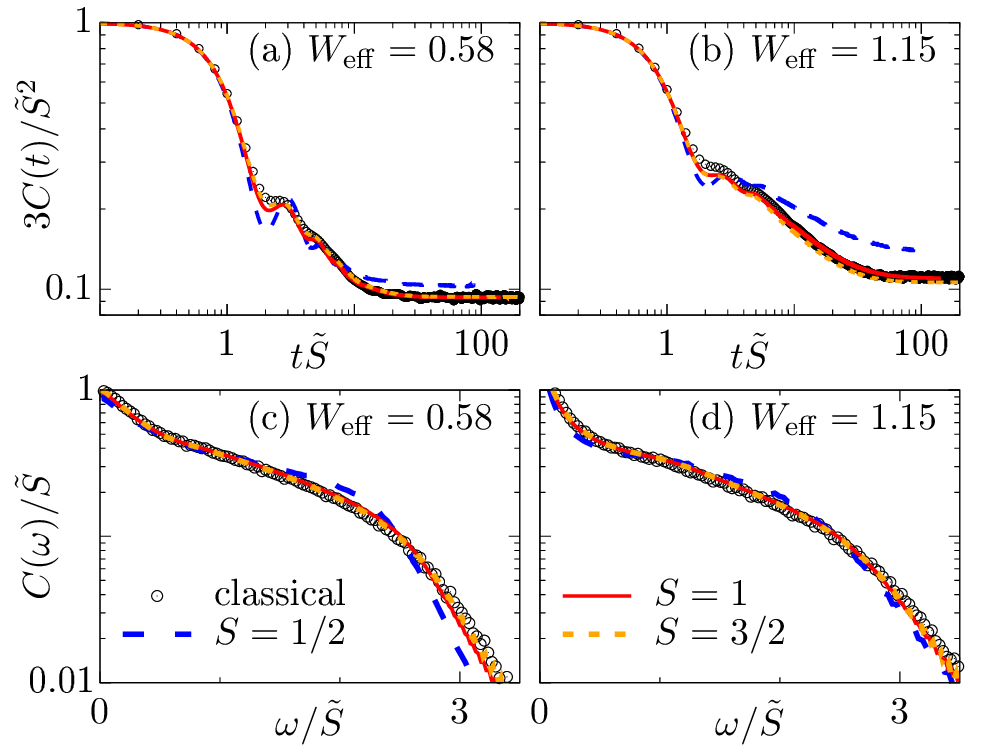}
 \caption{(Color online) 
  (a) and (b) Equal-site correlation function 
 $3C(t)/\widetilde{S}^2$ versus rescaled time $t\widetilde{S}$ for classical 
spins 
 (symbols) and $S = 1/2, 1, 3/2$ (dashed, solid, dotted) [see key in panels 
(c),(d)]. Data is shown for two different effective disorder strengths 
$W_\text{eff} \approx 0.58, 1.15$. (c) and (d) Corresponding spectral functions 
$C(\omega)/\widetilde{S}$ versus rescaled frequency $\omega/\widetilde{S}$. We 
have $L = 12$ 
in 
all cases.
}
 \label{Fig3}
\end{figure}

Let us start our numerical analysis with the case of small disorder. To this 
end, Fig.\ \ref{Fig3}~(a) shows the equal-site correlation function 
$C(t)$ for $W_\text{eff} \approx 0.58$ and a 
fixed system size $L = 12$. Similar to the case of $W = 0$ in 
Fig.\ \ref{Fig1}~(a), we observe that the curves for $S = 1,3/2$ and for 
classical spins are essentially indistinguishable from each other.
Moreover, while the long-time value of the $S = 1/2$ curve appears to be 
slightly higher compared to the curves of the other $S$, we show later in Sec.\ 
\ref{Sec::SA} that this finding is just a finite-size effect and becomes less 
pronounced for increasing $L$.
Remarkably, as shown in Fig.\ \ref{Fig3}~(b), this mapping between classical and 
quantum dynamics persists for $S = 1,3/2$ also at the slightly stronger 
disorder 
$W_\text{eff} 
\approx 1.15$. In contrast to the previous 
case in Fig.\ \ref{Fig3}~(a), however, we now find that the $S = 1/2$ model 
clearly deviates from the curves for $S \geq 1$ and exhibits distinctly 
slower dynamics, 
which is consistent with subdiffusive transport in this parameter 
regime. 

The corresponding spectral functions $C(\omega)$ for $W_\text{eff} \approx 
0.58, 1.15$ are presented in Figs.\ \ref{Fig3}~(c) and (d). Also for 
$C(\omega)$, the rescaling $\omega \to \omega/\widetilde{S}$ and 
$W \to W_\text{eff}$ leads to a convincing data collapse of all curves shown 
here. Interestingly, the clear deviations of the real-time $S = 1/2$ data 
observed in 
Fig.\ \ref{Fig3}~(b) are much less pronounced in the frequency 
representation [see Fig.\ 
\ref{Fig3}~(d)]. Moreover, we generally find that the  overall shape 
of $C(\omega)$ is rather similar compared to 
the disorder-free case in Fig.\ \ref{Fig1}~(b), except for the regime of very 
small frequencies $\omega \to 0$, where $C(\omega)$ starts to diverge 
upon increasing $W$. 
\begin{figure}[t]
 \centering
 \includegraphics[width=1\columnwidth]{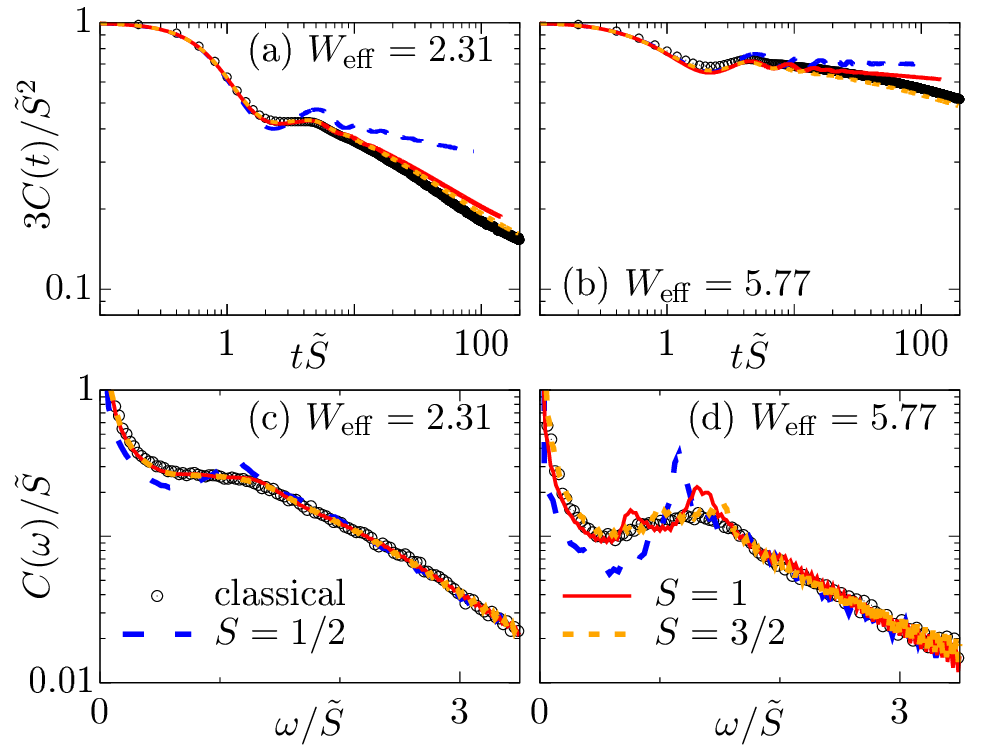}
 \caption{(Color online) 
Analogous data as in Fig.\ \ref{Fig3}, but now for stronger effective disorder 
$W_\text{eff} \approx 2.31, 5.77$.}
 \label{Fig4}
\end{figure}

To proceed, numerical results for the real-time correlation $C(t)$ at stronger 
disorder $W_\text{eff} = 2.31$ and 
$W_\text{eff} = 5.77$ are shown in Figs.\ \ref{Fig4}~(a) and (b). 
In the case of $W_\text{eff} = 2.31$, $C(t)$ now decays 
distinctly slower for all values of $S$, which 
might indicate that subdiffusion also occurs for larger spin quantum numbers $S 
= 1,3/2$. Moreover, compared to Fig.\ \ref{Fig3}~(b), we observe that not only 
the $S = 1/2$ data but also the $S = 1$ curve starts to deviate from the 
results for $S = 3/2$ and classical spins.
Furthermore, in the case of $W_\text{eff} = 5.77$ [Fig.\ \ref{Fig4}~(b)], we 
find that $C(t)$ becomes 
practically time-independent for $S = 1/2$, which signals the onset of 
many-body localization. In contrast, for $S = 1,3/2$ as well as 
classical spins, $C(t)$ clearly has a nonzero slope and continues to decay at 
long times (albeit this decay is very slow for $S = 1$ compared to $S = 
3/2$ and classical dynamics).   

Eventually, Figs.\ \ref{Fig4}~(c) and (d) show the respective spectral 
functions $C(\omega)$ at strong disorder. 
Compared to the cases of vanishing or small disorder in Figs.\ 
\ref{Fig1} and \ref{Fig3}, $C(\omega)$ now behaves qualitatively different and 
develops a long tail at large $\omega$. Nevertheless, at least for 
$W_\text{eff} = 2.31$, we find a reasonable data collapse of $C(\omega)$ for 
all values of $S$.
In Fig.\ \ref{Fig4}~(d), we moreover show that the onset of MBL in the 
spin-$1/2$ model reflects itself in a concave shape of $C(\omega)$ with a 
distinct peak at 
$\omega/\widetilde{S} \approx 1$, consistent with earlier results in Ref.\ 
\cite{Serbyn2017} (see also \cite{Prelovsek2017}). While such a pronounced 
feature is absent for $S = 3/2$ and classical spins, a small double-peak 
structure around $\omega/\widetilde{S} = 1$ emerges in the case of $S = 1$.

\subsubsection{Intermediate conclusion}

Given our numerical results in Figs.\ \ref{Fig3} and \ref{Fig4}, we conclude 
that for small values of disorder, the dynamics of $C(t)$ and $C(\omega)$ 
becomes almost independent of the quantum number $S$ when simulated at the 
same effective disorder strength $W_\text{eff}$.
In contrast, for stronger values of disorder, this mapping at 
least partially breaks down. Specifically, while the spin-$1/2$ model appears 
to be localized at $W_\text{eff} = 5.77$, this does not seem to be the case 
for $S = 1,3/2$ as well as classical spins. 
In particular, we have found a very good agreement 
between the dynamics for $S = 3/2$ and classical spins, at least for the 
values of $W_\text{eff}$ and the time scales considered in Figs.\ \ref{Fig3} 
and \ref{Fig4}. 

Generally, the dynamics 
of $C(t)$ and $C(\omega)$ in Figs.\ \ref{Fig3} and \ref{Fig4} is  
consistent with the behavior of the level-spacing distribution $\langle r 
\rangle$ discussed in 
the context of Fig.\ \ref{Fig2}~(b). Namely, for a fixed value of 
$W_\text{eff}$, $\langle r \rangle$ is still closer to the chaotic Wigner-Dyson 
value for a larger value of $S$. From the combination of Figs.\ 
\ref{Fig2}-\ref{Fig4}, one might speculate that MBL eventually 
occurs also in models with $S = 1$ and $S = 3/2$, but the critical disorder 
strength has to be even stronger than the largest $W_\text{eff} = 5.77$ 
considered by us. (Note that for $S = 3/2$, this 
$W_\text{eff}$ already corresponds to the very large bare disorder $W \approx 
11.18$, cf.\ 
Table \ref{Tab1}). In view of the small system sizes numerically accessible for 
$S = 1,3/2$, however, we here refrain from a more detailed analysis of the 
putative onset of MBL in these models. 

\subsubsection{Finite-size analysis and self-averaging}\label{Sec::SA}

While we have shown numerical result for $C(t)$ and $C(\omega)$ in 
Figs.\ \ref{Fig3} and \ref{Fig4} for a fixed system size $L = 12$, let 
us briefly comment also on the finite-size scaling of these quantities. Since  
the numerical calculation of larger system sizes for $S = 1,3/2$ becomes  
unfeasible very quickly, we here restrict ourselves to the cases of $S = 1/2$ 
and classical spins. 
\begin{figure}[tb]
    \centering
    \includegraphics[width=0.9\columnwidth]{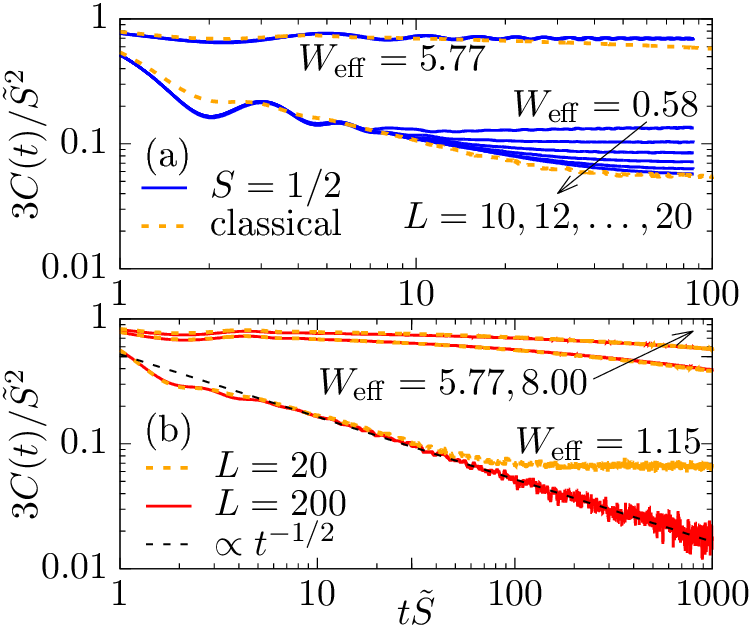}
    \caption{(Color online) (a) $3C(t)/\widetilde{S}^2$ for spin-$1/2$ 
chains with system sizes $L = 
10,12,\dots,20$ and effective 
disorder $W_\text{eff} \approx 0.58, 5.77$. As a comparison, we also depict 
corresponding $L = 20$ data for classical spin chains. 
(b) $3C(t)/\widetilde{S}^2$ in classical spin chains with $L 
= 20,200$ and disorder $W_\text{eff} \approx 1.15, 5.77, 8.00$ up to long time 
scales $t \leq 1000$. The dashed line indicates a diffusive power-law decay 
$\propto t^{-1/2}$. 
}
    \label{Fig5}
\end{figure}

To begin with, Fig.\ \ref{Fig5}~(a) shows the real-time correlation 
function $C(t)$ for spin-$1/2$ chains with different system sizes 
$L = 10,12,\dots,20$ and two different values of 
the effective disorder $W_\text{eff} \approx 0.58, 5.77$. As a comparison, we 
also depict corresponding data for classical chains with $L = 20$. On the one 
hand, for strong $W_\text{eff} = 5.77$, we observe that the spin-$1/2$ curves 
are essentially independent of $L$ for the time scales and systems sizes shown 
here.
On the other hand, for $W_\text{eff} \approx 0.58$, one finds that $C(t)$ 
remains converged 
up to times $t\widetilde{S} \lesssim 10$, while finite-size effects occur at 
later times and the long-time value scales as $C(t\to \infty) \propto 1/L$, as 
expected for the ergodic regime. 
Moreover, comparing the $L = 20$ 
curves for $S = 1/2$ and classical spins at this $W_\text{eff}$, we find that 
their long-time behaviors agree very well with each other.
Connecting to our earlier discussion in the context of Fig.\ \ref{Fig3}~(a), 
this observation confirms that the effective disorder $W_\text{eff}$, at least 
for small disorder, provides a useful mapping between classical and quantum 
dynamics with different $S$. 
 
Since it is possible to treat classical spin chains with significantly larger 
system sizes, Fig.\ \ref{Fig5}~(b) exemplarily shows numerical 
results for $L = 20$ and $L = 200$ up to rather long  
time scales $t \leq 1000$. (Note that this is still far from the maximum $L$ 
and $t$ values accessible \cite{Oganesyan2009, 
Steinigeweg2012}).
For weak $W_\text{eff} \approx 1.15$ and large $L = 200$, we 
observe a pronounced diffusive decay, 
$C(t) \propto t^{-1/2}$, which persists essentially over the 
entire time window depicted. Furthermore, for stronger $W_\text{eff} 
\approx 5.77, 8.00$, one finds that although the dynamics of $C(t)$ is very 
slow 
for $t \lesssim 100$, the clearly visible decay at longer times $t \gtrsim 500$ 
 is incompatible with localization. Comparing data for $L = 20$ and $L = 
200$, we moreover find that finite-size effects remain irrelevant at these 
large disorder values, even for 
the long time scales studied in Fig.\ \ref{Fig5}~(b). Our 
results indicate that while high-temperature spin transport in classical 
spin chains becomes strongly suppressed upon increasing disorder, a 
genuine MBL phase is absent in the classical model. These findings for the 
equal-site correlation function are consistent with earlier results from Ref.\ 
\cite{Oganesyan2009}, which has focused on energy transport instead.
\begin{figure}[tb]
 \centering
 \includegraphics[width = 0.85\columnwidth]{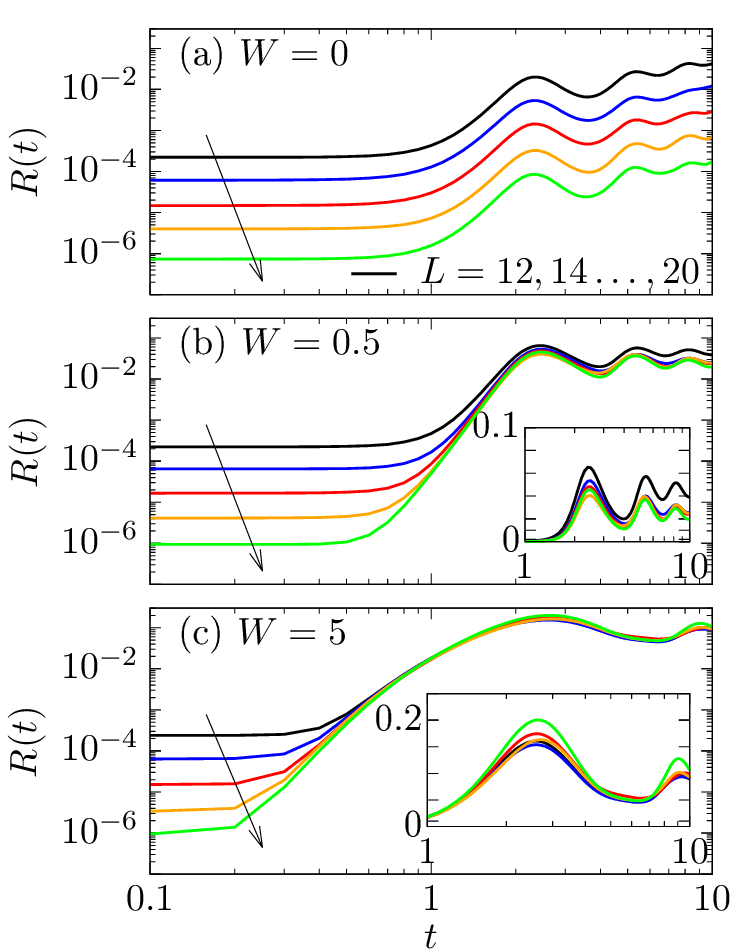}
 \caption{(Color online) Relative variance $R(t)$ of sample-to-sample 
fluctuations, cf.\ Eq.\ \eqref{Eq::Rt}, in a logarithmic plot for 
spin-$1/2$ chains 
with different system sizes $L = 12,14,\dots,20$ (arrows) and disorder 
strengths 
(a) $W = 0$, (b) $W = 0.5$ and (c) $W = 5$. The insets in (b) and (c) show  
close-ups of the time window $1 \leq t \leq 10$ and have a linear vertical 
axis. 
Note that the nonzero data of $R(t)$ in (a) is due to the typicality 
approximation which is used to obtain $C(t)$.}
 \label{Fig6}
\end{figure}

Next, let us discuss the self-averaging properties of $C(t)$. 
To this end, the relative variance $R(t)$ of sample-to-sample fluctuations 
[cf.\ Eq.\ 
\eqref{Eq::Rt}] is shown in Fig.\ \ref{Fig6} for spin-$1/2$ chains with
vanishing disorder $W = 0$, as well as weak and strong disorder $W = 0.5,5$. In 
particular, we restrict ourselves to short time scales $t \leq 10$, where 
$C(t)$ is free of trivial finite-size effects, cf.\ Fig.\ 
\ref{Fig5}~(a). 

On the one hand, for $W = 0$, $R(t)$ should in principle be strictly zero 
since there is no disorder. Note, however, that we calculate 
$C(t)$ by means of a typicality approach which relies 
on randomly drawn pure states and comprises a finite statistical error, cf.\ 
Eq.\ \eqref{Eq::Typ}. The nonzero data for $R(t)$ shown in Fig.\ 
\ref{Fig6}~(a) can therefore be interpreted as the accuracy of this typicality 
approximation. Consistent with our discussion in Sec.\ \ref{Sec::NumQD}, we 
find 
that $R(t)$ decreases exponentially upon increasing
$L$ for all times shown here. This demonstrates that typicality indeed provides 
an accurate numerical 
method to determine $C(t)$ and, in the spirit of Eq.\ \eqref{Eq::Rt}, $C(t)$ is 
``super'' self-averaging at $W = 0$. 
\begin{figure}[tb]
\centering
    \includegraphics[width=1\columnwidth]{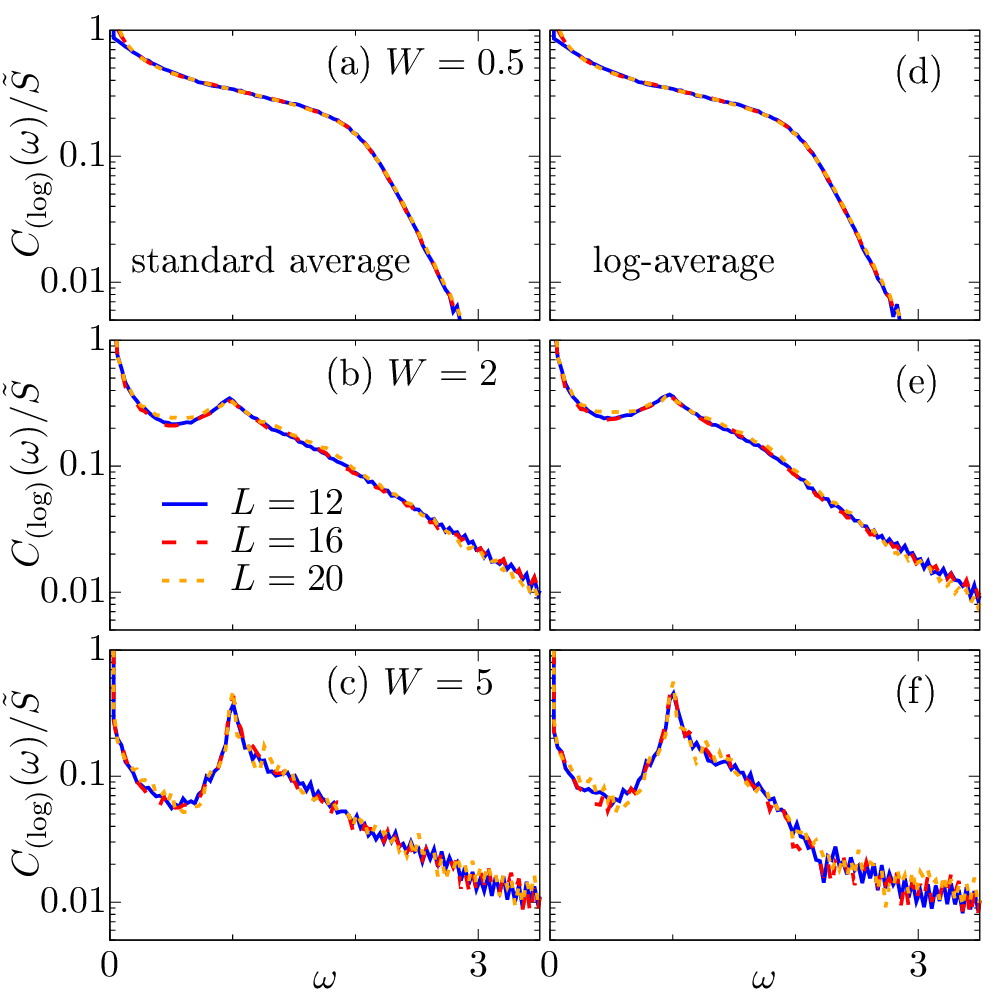}
    \caption{(Color online) Spectral functions for spin-$1/2$ chains with 
different system sizes $L = 12,16,20$ and different disorder strengths $W = 
0.5,2,5$. Panels (a) - (c) show results for the standard average 
$C(\omega)$, while panels (d) - (f) show the 
log-averaged version $C_\text{log}(\omega)$.}
    \label{Fig7}
\end{figure}

On the other hand, for finite disorder $W > 0$, the behavior of $R(t)$ is more 
complicated [see Figs.\ \ref{Fig6}~(b) and (c)]. For very short times, we 
find that the disorder apparently has no impact, such that 
$R(t)$ is dominated by the typicality contribution and decreases 
exponentially with $L$. In contrast, for longer times, this exponential scaling 
clearly breaks down and the curves for different $L$ are very similar to each 
other. While for $W = 0.5$ one might still argue that $R(t)$ slowly 
decreases with 
increasing $L$ [see also inset in Fig.\ \ref{Fig6}~(b) with 
linear axis], $R(t)$ 
appears to be independent of $L$ at strong $W = 5$ (except for 
residual statistical errors in our numerics). 
Consistent with 
recent results from Ref.\ \cite{Schiulaz2019}, this indicates 
that the equal-site correlation function $C(t)$ is self-averaging for short 
times $t 
\to 0$. At longer times, self-averaging is either much weaker (at small 
disorder), or might break down (at stronger disorder). Note that, since our 
conclusions are based on finite-size data with $L \leq 
20$, we cannot rule out that a power-law scaling $R(t) \propto L^{-\nu}$ 
eventually emerges for larger system sizes (especially if $\nu$ is small). 
    
To proceed, we discuss finite-size effects for the spectral function 
$C(\omega)$. In Figs.\ \ref{Fig7}~(a)-(c), $C(\omega)$ is shown for 
three different disorder values $W = 0.5,2,5$ and three different system sizes 
$L = 12,16,20$. 
Generally, one observes that $C(\omega)$ is 
essentially free of finite-size effects and the curves for different $L$ 
coincide very well with each other. The only exception to this finding is the 
case of weak disorder $W = 0.5$ and very small $\omega \to 0$ 
[Fig.\ \ref{Fig7}~(a)], where the curves do not collapse 
anymore (see also Ref.\ \cite{Serbyn2017} for similar results and further 
discussion).   

Eventually, we also study the log-averaged spectral function 
$C_\text{log}(\omega)$. In this context, let us note that Ref.\ 
\cite{Serbyn2017} investigated a very similar quantity 
and reported on a breakdown of self-averaging at large disorder. Specifically, 
Ref.\ \cite{Serbyn2017} found that the logarithmic average became strongly 
dependent on the system size $L$ at large $W$ (while it was $L$-independent for 
smaller $W$). As can be seen in Figs.\ \ref{Fig7}~(d)-(f), our numerical 
results for 
$C_\text{log}(\omega)$ do not confirm this finding. [Note that we have 
considered the 
same values of $W$ and $L$ as in the previous discussion of 
$C(\omega)$]. Namely, we find that the log-averaged spectral 
function $C_\text{log}(\omega)$ exhibits no notable dependence on $L$, both for 
weak as well as strong disorder. Moreover, 
$C_\text{log}(\omega)$ is very similar to the standard average $C(\omega)$ 
shown in Figs.\ \ref{Fig7}~(a)-(c). 
We explain the apparent discrepancy between our results and the findings from 
Ref.\ \cite{Serbyn2017} by the different ways the log-averages are defined (see 
Appendix \ref{Sec::AppLA} for more details.)


\section{Discussion}\label{Sec::Discussion}
\begin{figure}[b]
 \centering
 \includegraphics[width=0.85\columnwidth]{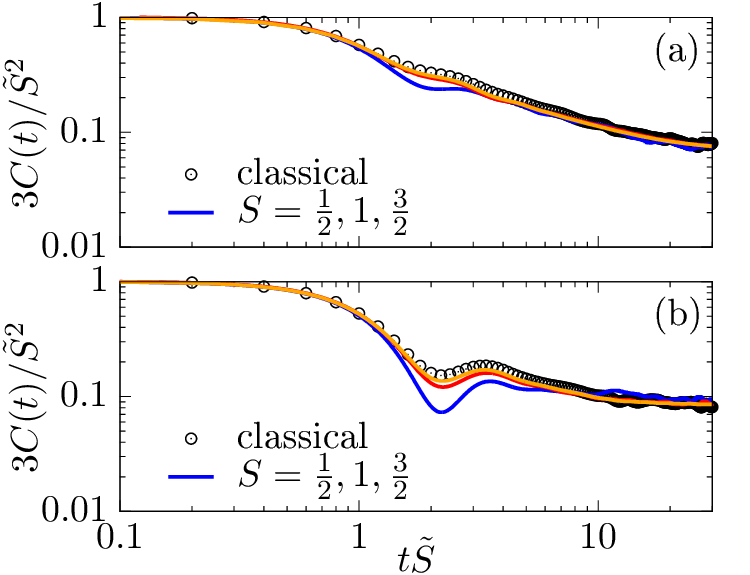}
 \caption{(Color online) (Color online) Equal-site correlation function 
$3C(t)/\widetilde{S}^2$ for spin quantum numbers $S = 1/2,1,3/2$ (curves) and 
classical spins (symbols) versus rescaled time
$t\widetilde{S}$. (a) $\Delta = 1.5$, $\Delta' = 0$, $L = 14$. (b) $\Delta = 
1$, 
$\Delta' = 1$, $L = 12$.}
 \label{Fig8}
\end{figure}

In conclusion, this work has shed light on the correspondence 
between quantum and classical 
dynamics, and moreover extended our understanding of the dynamics in disordered 
spin chains beyond the well-studied case of $S=1/2$. 
Specifically, we have compared the infinite-temperature dynamics of equal-site 
correlation functions $C(t)$ and their spectral functions $C(\omega)$ in 
disordered quantum and classical spin chains with $S = 1/2,1,3/2$.
Based on this comparison as well as by analyzing the statistics of 
energy-level spacings, we  
have shown that the putative many-body localization transition is shifted to 
larger 
and larger values of disorder upon increasing the spin 
quantum number $S$. 

Especially for vanishing or small values of disorder, we found that the 
dynamics of $C(t)$ and $C(\omega)$ becomes almost independent of $S$ and 
agrees with the classical result, upon introducing an effective 
disorder $W_\text{eff}$. Developing a better understanding and defining proper 
criteria where such 
a type of ``universality'' occurs, promises to be an interesting avenue of 
future 
research. This question is not only of conceptual relevance, but 
is also of practical importance if the dynamics of strongly correlated quantum 
systems can be obtained from a much less demanding simulation of a suitable 
classical spin model instead. In this context, it is also interesting to extend 
the comparison between quantum and classical dynamics to a wider class of 
observables such as, for instance, the full space-time profiles $\langle 
S_l^z(t) 
S_{l'}^z \rangle$ and their respective structure factors in momentum space. 

While the very good agreement 
between classical and quantum dynamics with $S = 3/2$ was found to persist even 
at very large values of $W_\text{eff}$ (given the system sizes and time 
scales available), the mapping between different quantum numbers $S$ at least 
partially 
breaks down at stronger disorder. Specifically, while $C(t)$ and 
$C(\omega)$ exhibit distinct signatures of MBL in the spin-$1/2$ model, the 
dynamics for $S = 1,3/2$ appear delocalized even at the 
strongest $W_\text{eff}$ considered by us.  
Clarifying the existence of a MBL 
transition and determining the exact critical disorder strength in models with 
$S > 1/2$ will require further 
numerical and analytical efforts in the future. 


\subsection*{Acknowledgments} 
We thank M. Serbyn for fruitful discussions. This work has been funded by the 
Deutsche Forschungsgemeinschaft (DFG) - Grants No.\ 397067869 (STE 2243/3-1), 
No.\ 355031190 - within the DFG Research Unit FOR 2692. 

\appendix

\section{Dynamics in clean models for additional model 
parameters}\label{Sec::App1}

In addition to the data presented in the context of Fig.\ \ref{Fig1}, let us 
briefly discuss the dynamics of $C(t)$ also for other disorder-free models.  
To this end, we extend the Hamiltonian \eqref{EqHamil} from the main text 
by considering an additional anisotropy in the $z$ direction and a next-nearest 
neighbor interaction, i.e., ${\cal H} = J\sum_{l=1}^L H_l$ now reads  
\begin{equation}
 H_l = S_l^x S_{l+1}^x + S_l^y S_{l+1}^y + \Delta S_l^z 
S_{l+1}^z + \Delta' S_l^z S_{l+2}^z\ . 
\end{equation}

In Fig.\ \ref{Fig8}~(a), $C(t)$ is shown for $\Delta 
= 1.5$ and $\Delta' = 0$. Note that also for this choice of $\Delta, \Delta'$, 
the spin-$1/2$ model remains integrable. In comparison to Fig.\ \ref{Fig1}, we 
find that the agreement between 
quantum dynamics at different $S$ and classical mechanics becomes even better 
in the anisotropic model. In particular, the oscillations of the $S = 1/2$ 
curve are much less pronounced.

To proceed, Fig.\ \ref{Fig8}~(b) shows $C(t)$ for
a finite next-nearest neighbor interaction $\Delta = 
\Delta' = 1$, which breaks the integrability. While one might expect that such 
a breaking of integrability improves the agreement between $S = 1/2$ and 
larger $S \geq 1$, the results in Fig.\ \ref{Fig8}~(b) do not confirm this 
expectation. 
Specifically, we observe a pronounced dip of the $S = 1/2$ data at 
$t\widetilde{S} 
\approx 2$, which is not present for $S = 1,3/2$ or classical spins.
\begin{figure}[tb]
\centering
\includegraphics[width=1\columnwidth]{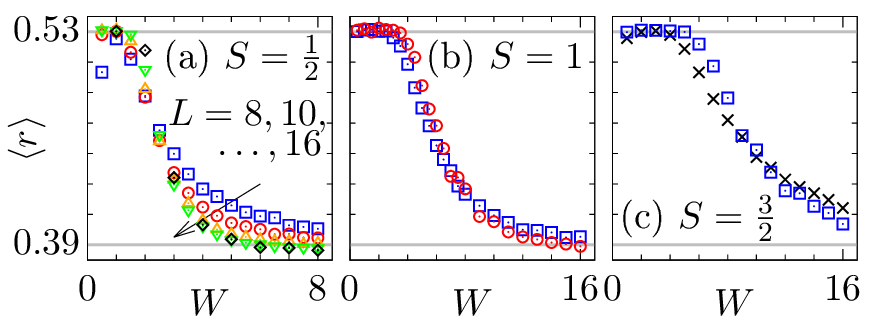}
\caption{(Color online) 
Finite-size scaling of the mean ratio 
$\langle r \rangle$ of adjacent level spacings for (a) $S = 1/2$ and $L = 
8,10,\dots,16$, (b) 
$S = 1$ and $L = 8,10$, 
and (c) $S = 3/2$ with $L = 6,8$.
}
\label{Fig9}
\end{figure}

\section{Finite-size effects of the level-spacing ratio 
$\langle r \rangle$}\label{Sec::FSLS}

In Fig.\ \ref{Fig9}, we show a finite size scaling of the mean ratio 
$\langle 
r \rangle$ of adjacent level spacings. As can be seen in Fig.\ 
\ref{Fig9}~(a) 
for the case of $S = 1/2$, the transition from Wigner-Dyson to Poissonian 
level statistics occurs more abruptly for increasing $L$ (see also 
\cite{Pal2010}).
Albeit the accessible system sizes are very small 
for $S = 1$ and $S = 3/2$, a similar behavior can also be found for these 
values of $S$ in Figs.\ \ref{Fig9}~(b) and (c). Therefore, one can expect
that the behavior of $\langle r \rangle$ discussed in the context of 
Fig.\ \ref{Fig2}~(b) can persist also for larger values of $L$. 

\section{Additional explanations and data on the log-averaged 
correlation functions}\label{Sec::AppLA}

Let us provide additional details on the log-averaged correlation functions 
and the discrepancy between our findings in Fig.\ \ref{Fig7} and the results 
from Ref.\ \cite{Serbyn2017}.
On the one hand, according to Eq.\ \eqref{Eq::CW}, we obtain 
$C_\text{log}(\omega)$ as the Fourier transform from the log-averaged real-time 
correlation function $C_\text{log}(t)$. Thus, explicitly writing 
$C_\text{log}(\omega)$ in terms of the eigenstates $\ket{n}$ and eigenvalues 
$E_n$ of the Hamiltonian ${\cal H}$, we have
\begin{align}\label{Eq::Long}
C_\text{log}(\omega) &= \int\limits_{-\infty}^\infty 
 \exp \left[\frac{1}{N_h} \sum_{h=1}^{N_h} 
\ln\left(\frac{1}{d}\sum_{n,m=1}^d e^{-i(E_m-E_n)t} \right. \right.  \nonumber 
\\  
 & \qquad \qquad  \left. \left. \times 
|\bra{m}S_l^z\ket{n}|^2\vphantom{\frac{1}{d}\sum_{n,m=1}^d}\right)\vphantom{
\frac{1}{N_h} \sum_{h=1}^{N_h} 
}\right]e^{i\omega t}\ \text{d}t\ , 
\end{align}
where the $\ket{n}$ and $E_n$ naturally depend on the specific disorder 
realization. Importantly, note that the logarithm in Eq.\ \eqref{Eq::Long} is 
outside the sum over the full Hilbert space. On the other hand, Ref.\ 
\cite{Serbyn2017} considered the average of logarithms of individual matrix 
elements, $\ln |\bra{m}S_l^z\ket{n}|^2$, which are then binned in 
respective energy windows. These two quantities are not necessarily the same 
and, in particular, the correlation function \eqref{Eq::Long} appears to be 
insensitive 
to the breakdown of self-averaging reported in \cite{Serbyn2017}. 

\begin{figure}[b]
 \centering
  \includegraphics[width = 0.85\columnwidth]{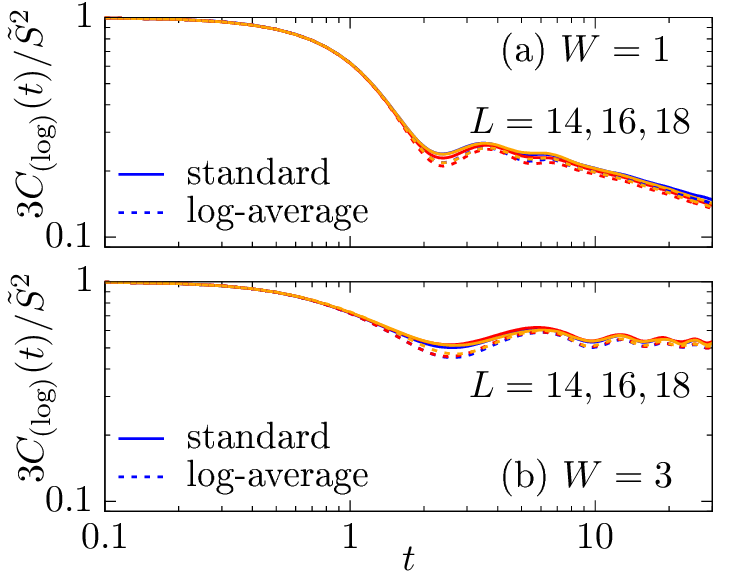}
  \caption{(Color online) Equal-site correlation function in spin-$1/2$ chains 
with $L = 14,16,18$ for (a) $W = 1$ and (b) $W = 3$. Solid lines are the 
standard average $C(t)$, while dashed lines indicate the log-averaged quantity 
$C_\text{log}(t)$.}
  \label{Fig10}
\end{figure}

Complementary to Fig.\ \ref{Fig7}, we present in Fig.\ \ref{Fig10} 
additional 
results for the log-averaged correlation $C_\text{log}(t)$ in real time. 
For the two disorder values $W = 1$ and $W = 3$ considered, we find that while 
the standard average $C(t)$ and the log-average $C_\text{log}(t)$ are slightly 
different from each other, their overall behavior is very similar (see also 
\cite{Agarwal2015}). Moreover, both $C(t)$ and $C_\text{log}(t)$ are almost 
independent of $L$, at least for the system sizes and time scales depicted. 
%


\end{document}